\documentclass[prl,twocolumn,floatfix,amsmath,amssymb]{revtex4}
\usepackage[ansinew]{inputenc}
\usepackage{graphicx}
\usepackage{pst-all}
\usepackage{color}
\usepackage{dcolumn}
\usepackage{amsmath}
\usepackage{amsthm}
\usepackage{bm}
\usepackage{layout}
\usepackage{float}
\usepackage{txfonts}
\usepackage{amsfonts}
\usepackage{amssymb}%
\usepackage[ansinew]{inputenc}
\usepackage{array}
\usepackage{multirow,bigdelim}
\usepackage{slashbox}
\usepackage{enumitem}

\begin{document}
\title{Dirac node lines in pure alkali earth metals }

\author{Ronghan Li$^{1}$}

\author{Xiyue Cheng$^{1}$}

\author{Ma Hui$^{1}$}

\author{Shoulong Wang$^{1}$}

\author{Dianzhong Li$^{1}$}

\author{Zhengyu Zhang$^{2}$}

\author{Yiyi Li$^{1}$}

\author{Xing-Qiu Chen$^{1}$}
\email[Corresponding author: ]{xingqiu.chen@imr.ac.cn}
\affiliation{$^{1}$Shenyang National Laboratory for Materials
Science, Institute of Metal Research, Chinese Academy of Science,
110016 Shenyang, Liaoning, China} \affiliation{$^{2}$ International
Center for Quantum Design of Functional Materials (ICQD), Hefei
National Laboratory for Physical Sciences at the Microscale (HFNL)
and Synergetic Innovation Center of Quantum Information and Quantum
Physics, University of Science and Technology of China, Hefei, Anhui
230026, P. R. China}
\date{\today}

\begin{abstract}
Beryllium is a simple alkali earth metal, but has been the target of
intensive studies for decades because of its unusual electron
behaviors at surfaces. Puzzling aspects include (i) severe
deviations from the  description of the nearly free electron
picture, (ii) anomalously large electron-phonon coupling effect, and
(iii) giant Friedal oscillations. The underlying origins for such
anomalous surface electron behaviors have been under active debate,
but with no consensus. Here, by means of first-principle
calculations, we discover that this pure metal system, surprisingly,
harbors the Dirac node line (DNL) that in turn helps to rationalize
many of the existing puzzles. The DNL is featured by a closed line
consisting of linear band crossings and its induced topological
surface band agrees well with previous photoemission spectroscopy
observation on Be (0001) surface. We further reveal that each of the
elemental alakali earth metals of Mg, Ca, and Sr also harbors the
DNL,  and speculate that the fascinating topological property of DNL
might naturally exist in other elemental metals as well.
\end{abstract}


\maketitle

Topological semimetals\cite{TSM} represent new types of quantum
matter, currently attracting widespread interest in condensed matter
physics and materials science. Compared with normal metals,
topological semimetals are distinct in two essential aspects: the
crossing points of the energy bands occur at the Fermi level, and
some of the crossing points consist of the monopoles in the lattice
momentum space. Topological semimetals can be classified into three
main categories, topological Dirac (TD)\cite{TD}, topological Weyl
(TW)\cite{TW} and Dirac node line (DNL)
semimetals\cite{TL-1,TL-2,TL-3}, respectively. In the former two
cases of TD and TW, the monopoles form isolated points in lattice
momentum space and novel surface states (i.e., surface Dirac cones
and Fermi-arc states) were observed or suggested, such as TD-type
Na$_{3}$Bi\cite{na3bi-1,na3bi-2,na3bi-3,na3bi-4} and
Cd$_3$As$_2$\cite{cdas1,cdas2,cdas3} and TW-type
TaAs-family\cite{taas1,taas2,taas3,taas4} and TW-type-II
WTe$_2$\cite{wte}.

In the third class of DNL, the crossings between energy bands form a
fully closed line nearly at the Fermi level in the lattice momentum
space, drastically different from the isolated Dirac (or Weyl)
points in the TD and TW. The projection of the Dirac node line into
a certain surface would result in a closed ring in which the
topological surface states (usually flat bands) can be expected to
appear due to the non-trivial topological property of its bulk
phase. According to the previous DNL modelings\cite{TL-1,TL-2}, the
band crossings occur at zero energy with a constraint chiral
symmetry, leading to the appearance of flat topologically protected
surface bands. However, in a real crystal the chiral symmetry of a
band structure is not exact, thereby suggesting that the DNL does
not generally occur at a constant energy and the DNL-induced
topological surface bands are not flat either. Recently, this type
of DNL states has been predicted in several cases of 3D carbon
graphene allotropes\cite{dnl1}, antiperovskite
Cu$_3$(Pd,Zn)N\cite{dnl2,dnl3}, Ca$_3$P$_2$\cite{dnl4},
LaN\cite{dnl5}, photonic crystals\cite{dnl6} and a hyperhoneycomb
lattice\cite{dnl7}, etc. But, all these DNL predictions have yet to
be experimentally verified. Here, we discover a new DNL state in
beryllium, the first known example in pure metals. We find that the
presence of the topologically protected (0001) surface states around
the $\bar{\Gamma}$ point originates from this DNL state.
Furthermore, realization of the existence of the DNL in principle
offers new opportunities and insights in  rationalizing the
long-standing problem\cite{be1,be2,be3,be4,be5,be6,be7,be8,be9} of
the pronounced surface states in Be.

The metal of beryllium, which crystallizes in the hcp structure (see
Fig. 1a), is a simple \emph{sp}-bonded metal. Be is unusual in three
aspects. Firstly, at its bulk phase it is almost a semimetal,
whereas its (0001) surface has the well-defined, intense and robust
surface electronic states\cite{be1,be2,be3}.Here, the surface
density of state at the Fermi level is almost five times higher than
that of its bulk phase\cite{be1,be8,be9}. The surface states of
simple metals were often interpreted within the framework of the
nearly-free-electron model\cite{be1}, however beryllium behaves
surprisingly far beyond this model. Secondly, the anomalous
interplanar expansion\cite{be7} as large as $>$ 4\% of the topmost
(0001) atomic layer has been observed, different from most other
pure metals which often show little relaxations or contractions when
cracking into surface. The underlying driving force for the lattice
expansion has again been attributed to unusual electronic states at
the surfaces\cite{sur-1,sur-2}. Thirdly, the large electron-phonon
coupling ($\lambda$ = 1.18 )\cite{be5} and giant Friedal
oscillations\cite{be6} have been found at its (0001) surface. This
fact along with the high density of states at the Fermi level was
suggested to be a candidate to show surface
superconductivity\cite{be4}.

\begin{figure}[hbt]
\centering
\includegraphics[width=0.45\textwidth]{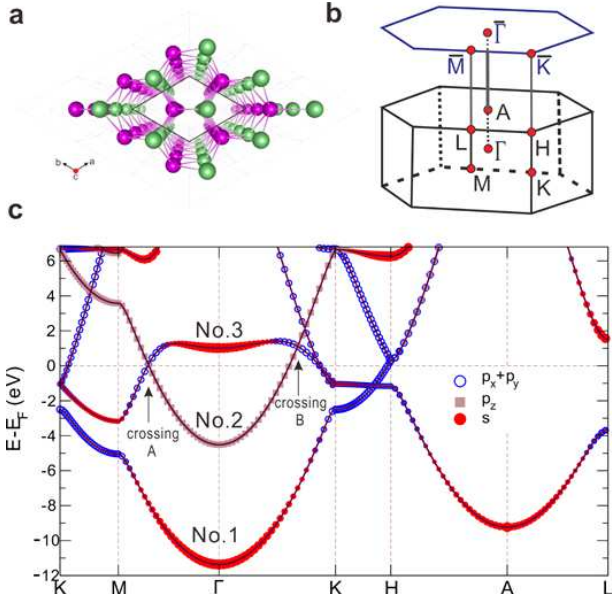}
\caption{\textbf{Lattice structure and electronic band structure of
hcp beryllium.} \textbf{a}, Crystal structure (here a 3
$\times$3$\times$5 supercell) (No.194 P63/mmc space group) with Be
atom at 2c site (1/3, 2/3, 1/4) in a perspective projection along
the c-axis. The layers consisting of green and pink atoms
alternatively stack along c-axis. \textbf{b}, Brillouin zone (BZ) of
the bulk phase and projected surface BZ of the (0001) surface,
\textbf{c} Calculated electronic band structure along the
high-symmetry points for hcp Be bulk phase.} \label{fig1}
\end{figure}

We start by representing the electronic structure of the bulk hcp Be
phase (Fig. 1a) at the ground state within the framework of density
functional theory (DFT) (Supplementary Materials). The derived
electronic band structure of the hcp Be is shown in Fig. 1c. It is
quite similar to previous calculations\cite{be-cal}. We also derive
the Fermi surface of its bulk phase in Fig. 2a, in a nice accordance
with the experimental measurement\cite{exp}. Its Fermi surface
clearly consists of two parts: a coronet-like shape of six-cornered
hole pockets on the k$_{z}$=0 plane, and a cigar-like shape of six
equivalent electronic surfaces along the H-K symmetry direction.
Nevertheless, a crucial feature was ignored in those previous
studies. That is the appearance of two clear band crossings featured
by a nearly linear dispersion around the Fermi level. One (crossing
A) exactly locates at the Fermi level in the M-$\Gamma$ direction
and the other one (crossing B) lies about 0.8 eV above the Fermi
level along the $\Gamma$-K direction. Actually, these two crossings
are induced by the band inversion. At the centre of the Brillouin
zone (BZ, see Fig. 1b), $\Gamma$, it can be seen that the band
\emph{s} $\rightarrow$ \emph{p}$_{z}$ inversion occurs between No.2
and No.3 bands due to the crystal field effect (Fig. 1c).
Remarkably, the band crossings indeed not only appear at these two
isolated A and B, but also form a circle-like closed line around the
$\Gamma$ point on the k$_z$=0 plane in BZ. This is the exact sign of
the DNL appearance. As elucidated in Fig. 2b, the band crossings
between No.2 and No.3 do not happen at the same energy, but show a
periodic wave-like closed curve upon the k vectors around the
centered $\Gamma$ point. In addition, this DNL's stability is highly
robust, protected by the inversion and time-reversal symmetries
without the spin-orbit coupling (SOC) effect. Because of the light
mass of Be, its SOC effect can be ignored.

\begin{figure}[hbt]
\centering
\includegraphics[width=0.45\textwidth]{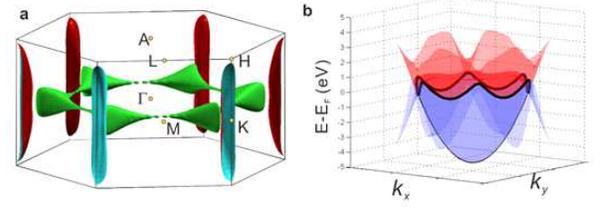}
\caption{\textbf{Fermi surface and the DNL in hcp Be. a}, The Fermi
surface in bulk hcp Be exhibits two-band features: a six-cornered
hole pockets around the K point in the $k_z$ = 0 plane and a six
column-like electronic pockets along the H-K direction. The hole
pockets are perpendicular to the electronic ones in the BZ.
\textbf{b}, Illustration of the DNL, highlighted by the two band
crossings between the No.2 and No.3 bands near the Fermi level on
the the $k_z$ = 0 plane for hcp Be.} \label{fig2}
\end{figure}

Accordingly, we further illustrate the topology of the DNL states in
Be. Since a Bl$\ddot{o}$ch Hamiltonian H(\textbf{\emph{k}}) is
invariant under the inversion (\emph{P}) and time-reversal
(\emph{T}) operations in a centrosymmetric system without SOC, it
can be inferred that \emph{T}$^2$=1 and (\emph{P$\cdot$T})$^2$=1.
The Berry phase $\gamma$ = $\oint_C$$A(\textbf{k})$d\textbf{k},
where $A(\textbf{k})$ is the Berry connection of
-$i$$\sum_n$$\langle$$u_n(\textbf{k})$$|$$\bigtriangledown_\textbf{k}$$u_n(\textbf{k})$$\rangle$
and $C$ is a closed loop in the momentum space,  can be used to
identify whether or not the DNL occurs\cite{dnl2,dnl3}. If the loop
C is pierced by a node line, the Berry phase $\gamma$ = $\pi$,
otherwise, $\gamma$ = 0. Within this relation, a $Z_2$ topological
invariant can be introduced as $\omega$($C$)=$\exp$
\emph{i}$\gamma$, which equals $\pm$1. Following the method proposed
by Fu and Kane\cite{TF}, the term of $\omega$($C$) can be expressed
as,
\begin{equation}
\omega(C_{ab}) = \xi_a\xi_b
\end{equation}
where $\xi_a$ is the product of the occupied band parities at $a$
point. Using a four parity-invariant points loop\cite{dnl3},
$C_{ab}-C_{cd}$ to define the boundary of the surface $S_{abcd}$,
the $Z_2$ invariant can count the number of node lines ($N(S)$)
which pierce the surface, as reads,
\begin{equation}
(-1)^{N(S_{abcd})} = \omega(C_{abcd}) = \xi_a\xi_b\xi_c\xi_d
\end{equation}
It implies that the DNL will occur around the inversion transition.
Additionally, we calculate the $Z_2$ indices at the eight
time-reversal invariant momentums (TRIMs) and obtaine that the
product is +1 at $\Gamma$ and -1 for all the other TRIMs, as shown
in TABLE \ref{tab:01}. It evidences the occurrence of a band
inversion at  $\Gamma$ and the DNL appearance around  $\Gamma$.
Since the relation is very similar to the topological invariants
($v_0$; $v_1$ $v_2$ $v_3$) of inversion-symmetric topological
insulator\cite{TF}. The $Z_2$ invariant in Be can be written as (1;
0 0 0).

\begin{table}[h]
\begin{tabular}{|c|c|c|c|c|} 
\hline
Momenta & $\Gamma$ (0, 0, 0) & A (0, 0, $\pi$) & M ($\pi$, 0, 0) $\times$3 & L ($\pi$, 0, $\pi$) $\times$3 \\
\hline
parities &+, + & +, -& +, -& +, - \\
\hline
produce &(+) &(-)&(-)&(-)\\
\hline
\end{tabular}
\caption{The parities of the occupied bands at the time-reversal
invariant momenta for hcp Be and the products of the parities
for the two occupied bands. }
\label{tab:01}
\end{table}

To further elucidate the topological feature in beryllium, we have
also constructed a 2$\times$2 Hamiltonian within the framework of
the low-energy \textbf{\emph{k$\cdot$p}} model in describing two
bands crossing, briefly. Using the $\mid$s$>$ and $\mid$p$>$ states
as the bases, respecting time-reversal, and D$_{6h}$ symmetries, the
model Hamiltonian around $\Gamma$ can be written as,
\begin{equation}
H_{\Gamma}(\textbf{k})=\epsilon(\textbf{k}) +
\begin{pmatrix}
M(\textbf{k})&B(\textbf{k}) \\
B^{\dagger}(\textbf{k})& -M(\textbf{k}) \\
\end{pmatrix}
\end{equation}
where
 \begin{equation}
\epsilon(\textbf{k}) = \epsilon_0+a_\parallel(k_x^{2}+k_y^{2})+a_{\perp}k_z^{2}
\end{equation}
 \begin{equation}
M(\textbf{k})=m_0+m_\parallel(k_x^{2}+k_y^{2})+m_{\perp}k_z^{2}
\end{equation}
\begin{equation}
B(\textbf{k})=B_0k_{z}
\end{equation}
and the eigenvalues of the model Hamiltonian are $E(\textbf{k})$ =
$\epsilon(\textbf{k})$$\pm$$\sqrt{M^2(\textbf{k})+B^2(\textbf{k})}$.
In this case, the gapless solutions of Equ. (3), at which the Dirac
node line would appear, can be yielded only when both
$M(\textbf{k})$ and $B(\textbf{k})$ equal zero. Within this
condition, on the k$_{z}$=0 plane the node line locates at
$\mid$\textbf{k}$\mid$ = $\pm$$\sqrt{-\frac{m_0}{m_\parallel}}$
which requires  $m_0$$m_\parallel$ is always smaller than zero. This
is also the condition of the band inversion occurrence. With other
words, when band inversion happens, in the lack of SOC, there always
exists a node line in the momentum space, which are the solutions of
$M(\textbf{k})$ = $B(\textbf{k})$ =0. Noted that the band inversion
only occurs at $\Gamma$ among all TRIMs for Be according to the
abovementioned topology analysis, a Dirac node line would occur
around the $\Gamma$ point in agreement with the calculated
electronic stricture.

\begin{figure}[hbt]
\centering
\includegraphics[width=0.45\textwidth]{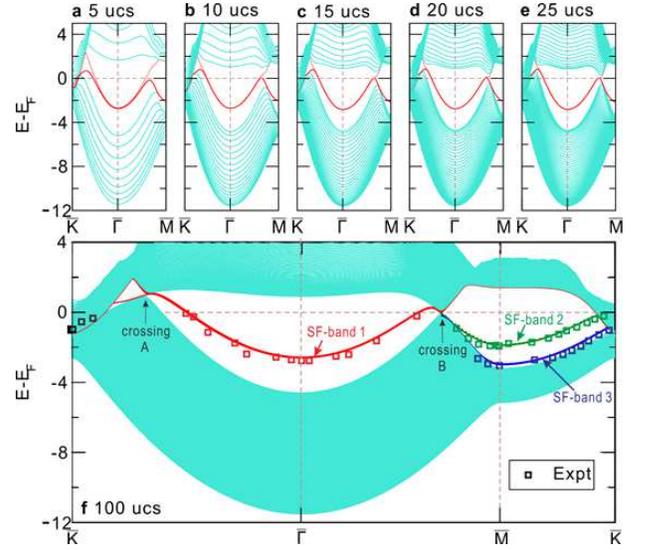}
\caption{\textbf{The evolution of thickness-dependent surface
electronic band structures of the Be (0001) surface. a} to
\textbf{e}, the derived surface electronic structure is shown as a
function of the slab thickness from 5 unit cells (ucs) to 25 ucs
along c-axis with an increasing of 5 ucs thickness of every image.
It can be seen that, with increasing the thickness, the surface
states (red curves) almost remains unchanged and the bands touch
more closely at both crossing A and B points, which correspond to
the band crossing points projected by the bulk energy bands.
\textbf{f}, Surface electronic band structure with a slab thickness
of 100 ucs, together with the experimentally measured data
(squares\cite{be1}).} \label{fig3}
\end{figure}

Using this Hamiltonian to describe the (0001) surface bands, the
topologically nontrivial surface states could be expected. Since the
(0001) surface is perpendicular to c-axis, we can use
-i$\partial_{z}$ instead of  k$_{z}$ according to the theory to the
linear order and then the (0001) surface Hamiltonian can be
expressed as follows,
\begin{equation}
H^{surf}_{\Gamma}(\textbf{k})=\epsilon_0+a_\parallel k^{2} +
\begin{pmatrix}
m_0+m_\parallel k^{2} & -iB_0\partial_{z} \\
iB_0\partial_{z}& -m_0-m_\parallel k^{2}\\
\end{pmatrix}
\end{equation}
Due to the Jackiw-Rebbi problem \cite{JR}, there will be a
topological boundary mode with the condition of $m_0$+$m_\parallel$
$k^{2}$ $<$ 0. This suggests the presence of the topologically
nontrivial surface states, which would nestle the projected Dirac
node ring. To clarify this point, we calculate the surface
electronic structures by varying the thickness of the slab as shown
in Fig. 3a-3f. As expected, the robust surface electronic bands
(SF-band 1 in Fig. 3f ) appear, when the slab's thickness is above
five unit cells along the c-axis. As shown in Fig. 3f, from the
crossing A to B point the topological surface bands are two-fold
degenerated within the projected Dirac node ring. It can be seen
that the topological surface bands along the
$\bar{K}$-$\bar{\Gamma}$-$\bar{H}$ direction disperse parabolically
from the lowest energy of -2.73 eV (expt\cite{be1,exp}: -2.80 eV and
-2.75 eV, referred to the Fermi level) at $\bar{\Gamma}$ , and then
cross the Fermi level at 47 \% (expt\cite{exp}: 49\%) of the
$\bar{\Gamma}$-$\bar{K}$ distance and 64\% (expt\cite{exp}: 58\%) of
the $\bar{\Gamma}$-$\bar{M}$ distance, in nice agreement with the
experimental findings (see squares in Fig. 3f) obtained by
ARPES\cite{be1,exp}. Particularly, it has been noted that the
surface bands around the $\bar{\Gamma}$ and $\bar{M}$  points are
highly different. The topologically protected SF-band 1 around the
$\bar{\Gamma}$ point are mainly comprised with $s$ and $p_z$-like
electronic states from the topmost atomic layer, reflecting well the
crucial feature of the band inversion between s and $p_z$ in its
bulk phase (Fig. 1c). This SF-band 1 is indeed half-filled when the
surface is electrically neutral. The other two surface bands around
$\bar{M}$ (SF-band 2 and 3 in Fig. 3f) are the topologically trivial
states, which are not correlated with the DNL state. The SF-band 2
is mainly composed with in-plane $p_{x,y}$-like electronic states
from the topmost surface atomic layer, whereas the SF-band 3 mainly
consists of the $p_{x,y}$-like electronic states from the second
outer atomic layer. Obviously, these SF-bands 2 and 3 are fully
occupied when the surface is electronic neutral.

\begin{figure}[hbt]
\centering
\includegraphics[width=0.45\textwidth]{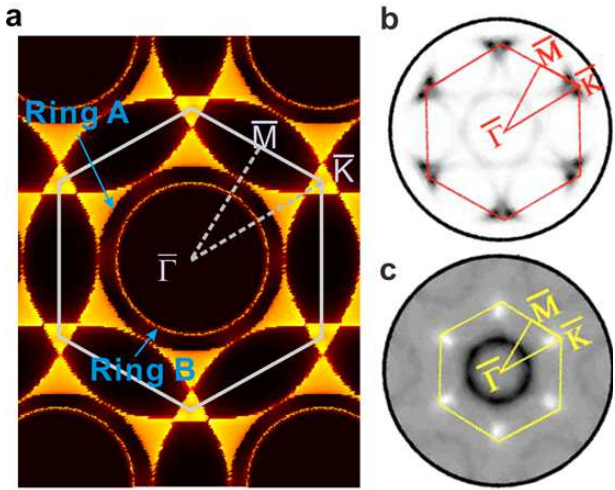}
\caption{\textbf{ Fermi surface of the Be (0001) surface.}
\textbf{a}, DFT-derived Fermi surface highlighting a surface state
of a fully closed circle around the centered $\bar{\Gamma}$ point in
the Brillouin zone. The other parts are the projections of bulk
bands. \textbf{b} and \textbf{c}, the experimentally measured Fermi
surface cuts with hv = 32.5 eV and 86 eV, respectively, showing the
difference between horizontal and vertical polarizations measured by
the ARPES experiments. The panels (\textbf{b}) and (\textbf{c}) are
directly adopted from the experimental report\cite{exp}.}
\label{fig4}
\end{figure}

We have further derived its Fermi surface of the (0001) surface in
Fig. 4a. Because the DNL exists on the $k_z$ = 0 plane in the BZ,
the DNL projection onto the (0001) surface shows a closed circle
(Ring A in Fig. 4a) surrounding $\bar{\Gamma}$ within which the
topologically protected surface bands (Ring B in Fig. 4a) appear.
Although the Ring B was already observed by ARPES\cite{exp} (Fig. 4b
and 4c), it was not interpreted  as the DNL-induced topologically
non-trivial bands. The understanding of the surface states (SF-band
1 in Fig. 3f and Ring B in Fig .4a) has also been a long-standing
question since the 1980s\cite{be1,be2,be3,be4,be5,be6,be7,be8,be9}.
It was suspected to be correlated with unusually large outward
relaxation of the topmost surface atoms, so-called \emph{p} to \emph{s} electron
demotion as well as the core-level
shifts\cite{be1,be2,be3,be7,be8,be9}. However, those proposed
mechanisms did not address the underlying physical origin for the
appearance and stability of the surface states\cite{be1}. Due to our
calculations, the SF-band 1 state is highly robust, no matter
whether the surface slab modeling are relaxed or not (Supplementary
Materials). Therefore, it is undoubtable that the SF-band 1 is
induced by the topological DNL feature in its bulk phase.
Looking back on the unusual physics properties of Be (0001) surface,
the puzzlings can be rationalized by the presence of topologically non-trivial DNL states.
The DNL induced topological surface bands would certainly behave far beyond the
nearly-free-electron model, and the nearly flat dispersion of them results in
very large electron density of state nearby the Fermi level of (0001)
surface, which would cause giant electron-phonon coupling effect and
Friedal oscillations. Additionally, we would like to mention that, as
early as in 1970, the Landau quantum oscillations of the magnetic
susceptibility has been observed when the magnetic field was set to
be parallel to the c-axis direction\cite{LQO}. This seems to be
another exciting sign to elaborate the DNL property in Be because
the magnetic field certainly breaks the time-reversal symmetry and
gap out the crossings to modify the Fermi surface.

Finally, we have also found the similar topological DNL feature in
Mg, Ca and Sr. Mg is isostructural and isovalent to Be and their
electronic band structures look highly similar to each other. In Mg
(which is similar to Be) there also exists a DNL around the Fermi
level and, the topologically protected non-trivial surface states
are also obtained on (0001) surface by surface calculations, which
are also in nice agreement with experiments(Supplementary
Materials). Metals of Ca and Sr, which are isovalent to Be and Mg,
crystallize in fcc structure with the truncated-octahedron Brillouin
zone. These two metals also exhibit the DNL-like topological
feature. On each hexagonal face of the Brillouin zone, the valence
and conduction bands cross each other to form a closed loop which
surrounds the center of the hexagonal face. Topological surface
states can also be observed clearly on their symmetrically
equivalent \{111\} surfaces. However, different from Be their
topological surface bands locate outside the projected DNLs because
the band inversion occurs outside the DNLs.

In summary, through first-principle calculations, we propose that 3D
topological DNL semimetal states can be obtained in pure alkali
earth metals of  beryllium, magnesium, calcium and strontium. The
anomalous topological surface bands are obtained on the (0001)
surface of Be (and Mg), and in nice agreement with previous
experimental observations. This fact confirms that the previous
observed anomalous electron behaviors on Be (0001) surface should be
induced by the topological non-trivial DNL state in its bulk and
rationalize a series long-standing puzzling physical properties on
its (0001) surface. The perfect agreement of physical properties
between theoretical predictions of DNL semimetals and experimental
observations on Be (0001) surface makes beryllium a good platform
for researching topological DNL semimetals. On the other hand, the
discovery of DNL states in alkali earth metals implies that this
fascinating topological property might also exist in other elemental
metals, supporting a new direction for designing topological DNL
semimetals.

\bigskip
\noindent {\bf Acknowledgments} We thank R. H. Hao for comments and
polishing in English on the manuscript. This study was supported by
the 'Hundred Talents Project' of the Chinese Academy of Sciences and
from the Major Research Plan (Grant No. 91226204), the Key Research
Program of Chinese Academy of Sciences (Grant No. KGZD-EW-T06), the
National Natural Science Foundation of China (Grant Nos. 51474202
and 51174188) and the Beijing Supercomputing Center of CAS
(including its Shenyang branch) as well as the high-performance
computational cluster in the Shenyang National University Science
and Technology Park.





\end{document}